\documentclass[prd,showpacs]{revtex4}
\usepackage{graphicx}
\usepackage{amsmath}

\providecommand{\U}[1]{\protect\rule{.1in}{.1in}}

\begin{document}

\title{\ Dilaton stabilization by massive fermion matter \bigskip }
\author{Alejandro Cabo$^{1}$, Matts Roos$^{2}$, Encieh
Erfani$^{3}$ \medskip}

\affiliation{$^{1}$\textit{Theoretical Physics Department, Instituto
de Cibern\'{e}tica, Matem\'{a}tica y F\'{\i}sica, Calle E, No. 309,
Vedado, La Habana, Cuba, }\\
$^{2}$ \textit{Physics Department, University of Helsinki, Helsinki, Finland,%
}\\
$^{3}$ \textit{Physikalisches Institut, Universit\"{a}t Bonn,
Nussallee 12, Bonn, Germany}}

\begin{abstract}
\noindent The study started in Ref. \cite{jcap} about the Dilaton mean field stabilization
thanks to the effective potential generated by the
existence of massive fermions, is here extended. Three loop corrections are evaluated in addition
 to the previously calculated two loop terms. The results indicate that the Dilaton vacuum
field tend to be fixed at a high value close to the Planck scale,
in accordance with the need for predicting Einstein gravity
from string theory. The mass of the Dilaton is evaluated to be
also a high value close to the Planck mass, which implies the absence of
Dilaton scalar signals in modern cosmological observations. These properties
arise when the fermion mass is chosen to be either at a lower bound
corresponding to the top quark mass, or alternatively, at a very much higher
value assumed to be in the grand unification energy range. One of the three
3-loop terms is exactly evaluated in terms of Master integrals. The other
two graphs are however evaluated in their leading logarithm correction in
the perturbative expansion. The calculation of the non leading logarithmic
contribution and the inclusion of higher loops terms could made more precise
the numerical estimates of the vacuum field value and masses, but seemingly
are expected not to change the qualitative behavior obtained. The validity
of the here employed Yukawa model approximation is argued for small value
of the fermion masses with respect to the Planck one. A correction to the
two loop calculation done  in the previous work is here underlined.

\bigskip

\noindent Accepted for publication in \textit{ Astrophysics and Space Science}.

\end{abstract}


\maketitle



\section{Introduction}

The Dilaton is an essential ingredient of superstring theory, and
constitutes a scalar field partner of the graviton \cite{GSW}. \ Therefore,
the background fields associated with the vacuum state of superstring theory
should involve this field in common with the metric in the basic action.
This is referred to as Dilaton gravity \cite{Ven,TV}. To the lowest level of
approximation the Dilaton is a free and massless scalar field with a special
kind of coupling to the matter fields. As a consequence of this coupling, a
time varying Dilaton field determines time-dependent coupling constants. In
order to overcome this difficulty the Dilaton should remain constant during
the present stage of evolution of the Universe. Moreover, unless the Dilaton
turns out to be very massive, its existence could lead to an observable
``Fifth force'' similar to the ones which are currently associated to the
observations of the Dark Matter. The constraints posed by current
experimental observations determine the lower bound on the mass of the
Dilaton to be of the order $m<10^{-12}\mathrm{GeV}$ \cite{bound} (but see
\cite{Polyakov} for an attempt to make a running Dilaton consistent with
late time cosmology). 

The Dilaton stabilization problem has been at the center of an intense
research activity in recent times because of its physical relevance. It
should be emphasized that the Dilaton is one of various scalar fields
appearing in the formulation of superstring theory in the low-energy limit.
The sizes and shapes of the extra spatial dimensions associated with
superstring theory are also leading to additional scalar fields, called
``moduli fields''. The stabilization of such moduli fields has been the
object of recent attention particularly in connection with Type IIB
superstring theory. The introduction of fluxes within the compactification
spaces has made it possible to stabilize various moduli fields \cite{GKP}.
Also, gaugino condensation \cite{gaugino} has been employed to stabilize the
Dilaton field in the context of heterotic superstring theory \cite{heterotic}
and in string gas cosmology \cite{Danos}.

It should be remarked that, since Dilaton stabilization has special
relevance for late time cosmology, there is motivation for finding
mechanisms which do not directly rest on the concrete assumptions defining
the nature of the extra dimensions. An additional motivation to search for
alternative Dilaton stabilization mechanisms comes from String Gas Cosmology
(SGC). The SGC \cite{BV,SGCrevs} is a model of early universe cosmology
which employs new degrees of freedom and symmetries of string theory, and
couples these elements with gravity and Dilaton fields into a classical
action background model. The Universe is considered to start as a compact
space containing a gas of strings. Since in string theory there is a maximal
temperature for a gas of closed strings, the initial state of the
cosmological evolution in SGC will be a phase of almost constant
temperature, the so called "Hagedorn phase". The SGC is able to define a
non-singular cosmology in which there is no starting Big Bang explosion. It
has been noted that the thermal fluctuations in a gas of closed strings in
the Hagedorn phase can justify the scale-invariant spectrum of cosmological
fluctuations observed in Nature \cite{NBV,BNPV2}, with a particular
prediction of a slight blue tilt for gravitational waves \cite{BNPV1}.
However, the consistency of the picture requires that the Dilaton field be
fixed during the Hagedorn phase. Therefore, in the SGC theory the Dilaton
needs to be fixed at very early times and at very late times.

Thus, clarifying the mechanisms of Dilaton field stabilization is an
important question in particle physics today. It is worth noting that the
universal type of coupling of the Dilaton to the matter fields not only
leads to an unwanted effect as the time-dependence of the coupling constants
but it also furnishes the possibility that quantum effects due to the
interaction of the Dilaton with matter might generate interesting
contributions to the effective potential of the Dilaton.

In a previous work
published in Ref. \cite{jcap}, we started to explore this question. The work
considered the cosmological periods when the additional spatial dimensions
of superstring theory were already stabilized and the study was done in the
framework of a four-dimensional field theory. The objective of study was
then the interaction of the Dilaton with massive fermions. Such masses can
be defined by fluxes about internal manifolds. In late time cosmology, the
masses could had been generated after supersymmetry breaking. In an
alternative early universe cosmology, one may consider thermally generated
fermion masses. In Ref. \cite{jcap} it was  considered a simple form for the Dilaton gravity
action in which a massive Dirac fermion term was added \cite{elizalde}. The
action was chosen in the Einstein frame, which does not show any Dilaton
field dependence in the kinetic terms for the fermions. On the other hand,
the fermion mass becomes a function of the Dilaton, involving a universal
exponential factor in Dilaton gravity \cite{Ven,TV}. The chosen action
described the low energy effective interaction of Super-Yang-Mills fermions
with the Dilaton field in superstring theory \cite{jcap}. The effective
potential for the Dilaton field was evaluated up to two loop corrections in
the small Dilaton radiative quantum field limit. \ That leads to a Yukawa
like interaction term which allows standard QFT calculations. A fixed value
of the cosmological scale factor was assumed. The outcome of the work was,
thanks to the appearing of logarithms in the loop calculations, that the
Dilaton field appeared in the result in powers multiplied by the exponential
factors of the field. This structure, in the one loop approximation clearly
indicated the spontaneous generation of vacuum mean value of the Dilaton
field.

Motivated by the dynamical generation of the Dilaton result in Ref. \cite
{jcap}, we here will address the evaluation of next corrections 3-loop terms
to the 2-loop evaluation of the effective potential for the Dilaton field.
The main issue to be \ explored is the possibility of the appearance in the
improved potential \ of the stabilizing effect which were in fact absent in
the second order correction, and which are suspected to be created by the
existence of massive matter upon the mean value of the Dilaton.

The results obtained, at least indicate, for the fermion mass being selected
at the $GUT$ or the $top$ quark mass scales, that the mean value of the
Dilaton field tends to be stabilized at a high value being close to the
Planck mass or the $GUT$ scale, respectively. Therefore, it is suggested
that the appearance of mass for matter in the course of the evolution of the
Universe can generate a stabilizing action on the vacuum expectation value
of the Dilaton field making it unobservable. This effect will tend to stop
the time evolution of the mean value, as it is convenient for String Theory
consistency.

The work also present an study of  the validity of the linear approximation
 of the Dilaton exponential factor which leads to the simpler effective Yukawa theory employed.
It follows  that for the two values of the large fermion masses assumed (the top quark and
GUT scale ones)  the approximation should  work well, after assuming that the low energy effective
effective action of string theory is the bare one in the renormalization of the model.
A brief resume about the procedure employed to estimate the Dilaton mass is also given.

\ We can point out, that in the process extending the work to
include higher loop corrections, we have noticed that in Ref.\
\cite{jcap} the kinetic term of the Dilaton Lagrangian was chosen
with a negative sign. This selection, although not changing the one
loop correction, led to a sign change of the 2-loop terms, which
suggested the existence of minima in the effective action argued in
Ref. \cite{jcap}. However, in spite of this non physical adopted
assumption in that work, the indication about the dynamical
generation in Ref.\cite{jcap} remained a valid one, because the
change in the metric did not affected the one-loop correction, the
basic quantity indicating the dynamical generation effect. The
present work corrects the result for the two loop terms, and
indicates that its place in the stabilizing effect over the Dilaton
field is played by higher order contributions.

The paper proceeds as follows: In Section II, the notation and basic
formulation are given. Section III presents the elements of the three loops
evaluation of the effective potential. \ Section IV \ discuss
the results of the calculation. \ In  Section V the results
are resumed and commented.  Finally, Appendix A presents the investigation of the
 Yukawa model approximation and the review on the scheme for evaluating the
 generated Dilaton mass.

\section{The Dilaton action and generating functional}

Let us consider a model of the Dilaton field interacting with fermion matter
in the form
\begin{eqnarray}
S\, &=&\,\int d^4x\sqrt{-g(x)}{\LARGE (}\frac 1{2\kappa ^2}g^{\mu \nu
}(x)\partial _\mu \phi ^r(x)\partial _\nu \phi ^r(x)+\overline{\Psi }(x)(i%
\frac{g^{\mu \nu }\gamma _\mu \overleftrightarrow{\partial }_\nu }2-m)\Psi
(x) \nonumber \\
&&\,-\overline{\Psi }(x)g_{Y\text{ }}^{*}\phi ^r(x)\Psi (x)+j(x)\phi ^r(x)+%
\overline{\Psi }(x)\eta (x)\text{ \ +}\overline{\eta }(x)\Psi (x){\LARGE )},
\\
m &=&\exp (\alpha ^{*}\text{ }\phi )m_f, \\
g_Y^{*} &=&\alpha ^{*}\text{ }m, \\
\alpha ^{*} &=&-\frac 34, \\
\text{ \ \ \ \ }x^\mu &=&(x^0,x^1,x^2,x^3),\text{ \ \ \ }\overleftrightarrow{%
\partial }=\overrightarrow{\partial }-\overleftarrow{\partial },\text{ \ }%
\left\{ \gamma _\mu ,\gamma _\nu \right\} =2g_{\mu \nu }(x), \\
g_{\mu \nu }(x) &=&\left(
\begin{array}{llll}
1 & 0 & 0 & 0 \\
0 & -1 & 0 & 0 \\
0 & 0 & -1 & 0 \\
0 & 0 & 0 & -1
\end{array}
\right) ,\text{ \ \ }\sqrt{-g(x)}=1.
\end{eqnarray}
That is, we are considering \ the Dilaton field \ interacting with a
massive fermion in the Einstein frame, in which the metric $g_{\mu
\nu }$ has been approximated by the Minkowski metric in order to
simplify the evaluation. The gravitational constant is here
explicitly introduced, and natural units are employed for the
distances and mass. The vacuum value of the Dilaton field is named
as $\phi $ and its radiative part is called $\phi ^r.$ Note that we
are assuming the radiative part is small in order to retain only the
first term in the \ expansion of the exponential. This is the Yukawa
approximation which is here employed.  In appendix A it is argued that it can be a
 a good approximation for the two values of the large fermion masses considered here :
 the top quark mass and a GUT scale one, assumed that the the ratio between the fermion mass
 and the Planck one is  very much smaller than one. All the results will be functions of the vacuum field $\phi$
 and the fermion mass $m$.

The parameter defining the Dilaton field dependent exponential, the Planck
length $\kappa =$ $l_P$ and mass $m_P$ are defined by the expressions
\begin{eqnarray}
\kappa ^2 &=&\frac{8\pi G
\rlap{\protect\rule[1.1ex]{.325em}{.1ex}}h%
}{c^3}, \\
\kappa &=&l_P=\frac 1{m_P}=8.10009\times 10^{-33}\ \text{cm}, \\
G &=&6.67\times 10^{-8}\ \text{cm}^3\ \text{g}^{-1}\ \text{s}^{-2}, \\
\hbar &=&1.05457\times 10^{-27}\ \text{cm}^2\ \text{g}\ \text{s}^{-1}, \\
c &=&2.9979245800\times 10^{10}\ \text{cm}\ \text{s}^{-1}.
\end{eqnarray}

In the above formula for the action, the coordinates and times are measured
in cm, the masses $m$ in the natural unit cm$^{-1}$ and the Dilaton field is
dimensionless.

Starting from the classical action, we will consider a 3-loop correction to
the effective action, assuming a homogenous and time independent value of
the Dilaton mean field $\phi $ as
\begin{equation}
\frac{\Gamma [\phi ]}{V^{(4)}}=-V^{eff}(\phi ),
\end{equation}
where $V^{(4)}$ is the four dimensional volume. In order to eliminate the
explicit appearance of the gravitational constant from the diagram technique
for evaluating the effective action, we could absorb it by redefining the
Dilaton field value and the $\alpha ^{*}$ constant as
\begin{eqnarray}
\ \varphi &=&\phi /\kappa , \\
\alpha &=&\alpha ^{*}\kappa =-\frac 34\kappa , \\
g_Y &=&g_Y^{*}\kappa .
\end{eqnarray}
After these changes, the above written classical action $S,$ to be used for
generating the Feynman expansion can be expressed as follows
\begin{eqnarray}
S\,[\overline{\Psi },\Psi ,\varphi ^r,\varphi ] &=&\,\int d^4x\text{ }%
{\Large (}\frac 12g^{\mu \nu }(x)\partial _\mu \varphi ^r(x)\partial _\nu
\varphi ^r(x)+\overline{\Psi }(x)(i\frac{g^{\mu \nu }\gamma _\mu
\overleftrightarrow{\partial }_\nu }2-m)\Psi (x)  \nonumber \\
&&-\overline{\Psi }(x)g_{Y\text{ }}\varphi ^r(x)\Psi (x)+j(x)(\text{ }%
\varphi +\varphi ^r(x))+\overline{\Psi }(x)\text{ }\eta (x)\text{ \ +}%
\overline{\eta }(x)\text{ }\Psi (x){\Large )}.
\end{eqnarray}

The expansion is considered in $\ d=4-2\epsilon $ dimensions for
implementing dimensional regularization scheme. Accordingly, the coupling
constant $g_{Y\text{ }}$ should be modified \ by the introduction of the
regularization scale parameter $\mu $ as \ follows \ \
\[
g_{Y\text{ }}^2=\mu ^{2\epsilon }(g_{Y\text{ }}^0)^2,
\]
where $g_{Y\text{ }}^0$ is the usual coupling constant in four dimensions.

\subsection{The generating functional and \ the effective action}

In this subsection, for the sake of definiteness, we will sketch the main
expressions defining the perturbative calculation to be considered \ in what
follows. \ \ The generating functional of the Green functions $Z$ , its
connected part $W$ \ and the mean field values will defined by the formulae

\ \ \ \ \ \ \ \ \ \ \ \ \ \ \ \ \ \ \ \ \ \ \ \ \ \ \ \ \
\begin{eqnarray}
Z[\overline{\eta },\eta ,j] &=&\int \mathcal{D}\overline{\Psi }\mathcal{D}%
\Psi \mathcal{D}\varphi ^r\exp (i\text{ }S\,[\overline{\Psi },\Psi ,\varphi
^r,\varphi ]), \\
W[\overline{\eta },\eta ,j] &=&\frac 1i\log Z[\overline{\eta },\eta ,j], \\
\frac{\delta \text{ }W}{i\text{ }\delta j(x)} &=&\varphi +\langle \varphi
^r(x)\rangle , \\
\frac{\delta \text{ }W}{i\text{ }\delta \overline{\eta }(x)} &=&\langle \Psi
(x)\rangle , \\
\frac{\delta \text{ }W}{-i\text{ }\delta \eta (x)} &=&\langle \overline{\Psi
}(x)\rangle .
\end{eqnarray}

Note that the mean Dilaton field $\varphi $ is considered as homogeneous and
the mean value of the radiative part $\langle \varphi ^r(x)\rangle $ will be
assumed to vanish when the sources are zero. \ The effective action is
defined as the Legendre transform of $Z$ \ depending on the mean field
values as:

\begin{eqnarray}
\Gamma [\langle \Psi \rangle ,\langle \overline{\Psi }\rangle ,\varphi
+\langle \varphi ^r\rangle ] &=&\frac 1i\log Z[\overline{\eta },\eta
,j]-\int dx{\Large [}j(x)(\text{ }\varphi +\langle \varphi ^r(x)\rangle
)+\langle \overline{\Psi }(x)\rangle \text{ }\eta (x)\text{ \ +}\overline{%
\eta }(x)\text{ }\langle \Psi (x)\rangle {\Large \ ]}, \\
\frac{\delta \text{ }\Gamma }{\delta \langle \varphi ^r(x)\rangle } &=&-j(x),
\\
\frac{\delta \text{ }\Gamma }{\delta \langle \overline{\Psi }(x)\rangle }
&=&-\eta (x), \\
\frac{\delta \text{ }\Gamma }{\delta \langle \Psi (x)\rangle } &=&\overline{%
\eta }(x).
\end{eqnarray}

\begin{figure}[h]
\begin{center}
\hspace*{-0.4cm} \includegraphics[width=7.5cm]{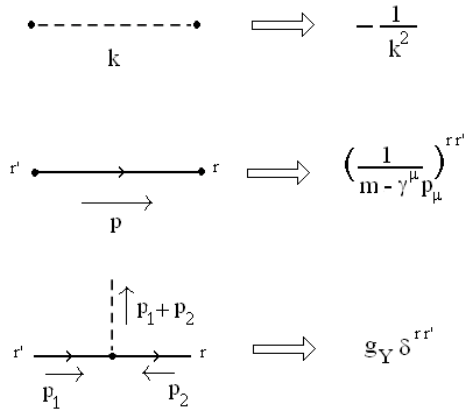}
\end{center}
\caption{ The figure illustrates the Feynman rules for the particular Yukawa
model approximation adopted for the Dilaton action}
\label{figura0}
\end{figure}

The expression for $Z,$ after writing the Yukawa vertex part of the
Lagrangian in terms of the functional derivatives over the sources and
integrating the gaussian functional integral that remains, leads to the Wick
expansion \ formula:
\begin{eqnarray}
Z[\overline{\eta },\eta ,j] &=&\exp {\large [}i\int dx\text{ }g_Y\frac
\delta {i\delta j(x)}\frac \delta {-i\delta \eta (x)}\frac \delta {i\delta
\overline{\eta }(x)}{\large ]\times }  \nonumber \\
&&\exp \left[ \int dx\text{ }dy\text{ }{\LARGE (}\overline{\eta }%
(x)S(x-y)\eta (y)+\frac 12j(x)D(x-y)j(y){\LARGE )}\right] , \\
S(x-y) &=&\int \frac{dp^d}{(2\pi )^d}\frac{\exp (-i\text{ }p.(x-y))}{%
m-\gamma ^\mu p_\mu }, \\
D(x-y) &=&\int \frac{dk^d}{(2\pi )^d}\frac{\exp (-i\text{ }k.(x-y))}{%
-(k^2-i\epsilon )},
\end{eqnarray}
in which $\ S$ and $D$ are the fermion and Dilaton free propagators,
respectively. \ The notation for fermions and scalar field related
quantities, and the definition of the Feynman rules for the \ \
generation of the analytic expressions for the various contributions
are exactly the ones described in Ref. \cite{muta}, for the cases of
scalar and fermion fields. \ \ Specifically, for the momentum space
rules, the propagators and \ the only existing vertex are
graphically illustrated in figure \ref{figura0}. \

\section{Effective potential evaluation}

\begin{figure}[h]
\begin{center}
\hspace*{-0.4cm} \includegraphics[width=7.5cm]{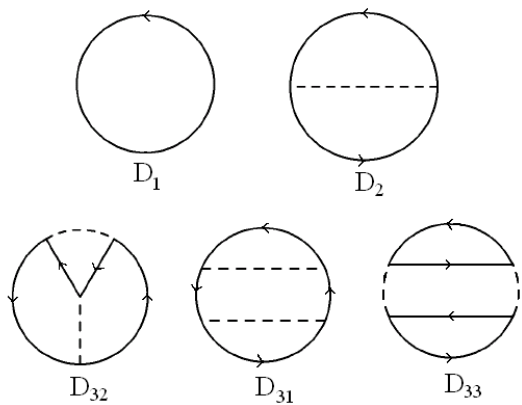}
\end{center}
\caption{The one, two and three loops Feynman diagrams considered in this
work. The one and two loop corrections $D_1$ and $D_2$ are exactly
calculated. In the case of the three loops terms, the $D_{32}$ is completely
evaluated in terms of the listed Master integrals in Ref. \protect\cite
{schavuor}. The $D_{31}$ and $D_{33}$ were determined only in their leading
logarithm correction. }
\label{figura1}
\end{figure}

Let us present in this section the evaluations of the effective potential
for the Dilaton field following after employing the perturbative expansion
described in the past section. \ The diagrams to be considered are depicted
in Fig. \ref{figura1}. They include up two three loop corrections. \ The
contributions will be exactly evaluated for the one and two loops. \ In
addition, the three loop term $D_{32}$ also can be analytically calculated
in terms of Master integrals. \ However, the three loop diagrams $D_{31}$ and $%
D_{33}$ \ are here determined only in their leading terms of order $\log
\left( \frac m\mu \right) ^3$ . We expect to be able in evaluating the non
leading corrections (lower powers of $\ log\left( \frac m\mu \right) $) in
extending the work.\ Let us discuss the results for each diagram in various
subsections below.

\subsection{One loop term $D_1$}

\ \ \ \ The analytic expression for the one loop diagram $\ D_1$ and its
derivative over $m^2$ have the forms

\begin{eqnarray}
\Gamma ^{(1)} &=&V^{(d)}\int \frac{dp^d}{(2\pi )^di}Tr\log (m^2-p^2), \\
\frac d{d\text{ }m^2}\Gamma ^{(1)} &=&4V^{(d)}\int \frac{dp^d}{(2\pi )^di}%
\frac 1{m^2-p^2}.
\end{eqnarray}
\ The result for the momentum integral entering in the derivative of $\ \Gamma
^{(1)}$ over $m^2,$ after divided by $\mu ^{2\epsilon }$ $V^{(d)}$ (in order
to define a 4-dimensional energy density) \ and integrated over $m^2$,
allows to write for the one loop effective action density (See Ref. \cite
{schavuor})
\begin{equation}
\gamma _1(m,\epsilon ,\mu )\equiv \frac{\Gamma ^{(1)}}{\mu
^{2\epsilon }V^{(d)}}=m^4(\frac m\mu )^{-2\epsilon }\frac{8\pi
^{2-\epsilon }}{(2\pi )^{4-2\epsilon }}\Gamma ( -1+\epsilon ).
 \end{equation}

After employing the minimal substraction (MS) scheme, that is,
getting the finite part
by eliminating the pure pole part in $\epsilon $ the Laurent expansion \ of $%
\gamma (m,\epsilon )$ and taking the limit $\epsilon \rightarrow 0,$ \ the
one loop contribution to the effective action density as a function of $\ m$
and $\mu $ \ \ becomes

\begin{equation}
\gamma _1(m,\mu )=0.0506606m^4\left( 2.\log \left( \frac m\mu \right)
-2.95381\right) .
\end{equation}

Note that the negative of this term, which gives the one loop
effective potential \ leads to a the dynamical generation of the
Dilaton field for positive values of $\alpha ^{*}$ $\phi $ as
follows from $log(m)=log(m_f)$ $+\alpha ^{*}$ $\phi .$ This was the
effect which motivated the study started in \ Ref. \cite{jcap}.

\subsection{ Two loop term $D_2$}

\ \ \ For the \ two loop contribution \ $D_2$ \ the analytic expression is

\begin{eqnarray}
\gamma _2(m,\epsilon ,\mu ) &\equiv &\frac{\Gamma ^{(2)}}{\mu ^{2\epsilon }V^{(d)}}%
=\frac 12(g_{Y\text{ }}^0)^2\text{ }\int \frac{dp_1^d}{(2\pi )^di}\frac{%
dp_2^d}{(2\pi )^di}\frac{4(m^2+p_1.p_2)}{(m^2-p_1^2)(m^2-p_2^2)(p_1-p_2)^2}
\nonumber \\
&=&\frac 12(g_{Y\text{ }}^0)^2m^{2d-4}\int \frac{dq_1^d}{(2\pi )^di}\frac{%
dq_2^d}{(2\pi )^di}\frac{4(1+q_1.q_2)}{(1-q_1^2)(1-q_2^2)(q_1-q_2)^2}
\nonumber \\
&=&2(g_{Y\text{ }}^0)^2m^4m^{-4\epsilon }{\Huge (}2\int \frac{dq_1^d}{(2\pi
)^di}\frac{dq_2^d}{(2\pi )^di}\frac 1{(1-q_1^2)(1-q_2^2)(q_1-q_2)^2}-
\nonumber \\
&&-\frac 12(\int \frac{dq_1^d}{(2\pi )^di}\frac 1{(1-q_1^2)})^2{\Huge )},
\end{eqnarray}
where the identity $q_1.q_2=\frac 12(q_1^2-1+q_{_2}{}^2-1)+1-\frac
12(q_1-q_2)^2$ has been used. \ The two momentum integrals appearing in the
last line are the simplest Master integrals for scalar fields as listed in
Ref. \cite{schavuor}. The results for them in that reference are:
\begin{eqnarray}
\int \frac{dq_1^d}{(2\pi )^di}\frac{dq_2^d}{(2\pi )^di}\frac
1{(1-q_1^2)(1-q_2^2)(q_1-q_2)^2} &=&\frac{(d-2)(\pi )^d\Gamma \left( 1-\frac
d2\right) ^2}{2(d-3)(2\pi )^{2d}}, \\
\int \frac{dq_1^d}{(2\pi )^di}\frac 1{(1-q_1^2)} &=&\frac{(\pi )^{\frac
d2}\Gamma \left( 1-\frac d2\right) }{(2\pi )^d}.
\end{eqnarray}

They allow to write for the regularized two loop effective action density
the expression

\begin{equation}
\gamma _2(m,\epsilon ,\mu )=-m^4(\frac m\mu )^{-4\epsilon }\frac{2(g_{Y\text{
}}^0)^2(\pi )^d}{(2\pi )^{2d}}(-\frac{d-2}{d-3}+\frac 12)\Gamma \left(
1-\frac d2\right) ^2.
\end{equation}

Expanding in Laurent series in $\epsilon $ and disregarding the \ pole part
\ in the limit $\epsilon \rightarrow 0$, leads to the two loop perturbative
contribution to the effective action
\begin{equation}
\gamma _2(m,\mu )=\text{ \ }0.0000200507\text{(}g_{Y\text{ }}^0)^2m^4\text{ }%
{\Huge (}48.\log ^2\left( \frac m\mu \right) -173.783\log \frac m\mu \text{
\ }+183.83\ {\Huge )}.
\end{equation}

As it was noticed in the Introduction, in Ref. \cite{jcap} it was employed
an inappropriate negative kinetic term for the Dilaton field. This change,
although not affecting the one fermion loop contribution, which is not
altered by the sign of the boson propagator, \ drastically modified the sign
of the two loop term which linearly depends on the Dilaton propagator. \ In
the \ previous evaluation, the two loop terms determined the existence of
minima for the Dilaton potential. Therefore, the consequence of the change
in sign fixed by the here consideration of the correct positive kinetic
energy term, should be investigated in connection with the existence of
stabilizing minima for the scalar field. This circumstance determined the
motivation for the new three loop corrections considered in this work.

\subsection{Three loops terms}

Let us now consider the three loop terms.

\subsubsection{Diagram \ $D_{32}$}

\ \ \ The $D_{32}$ term is the only of the 3-loops diagrams which is not
composed of two fermion or boson self energy insertions connected in series.
For the $D_{31}$ and $D_{33}$ cases we had difficulties in reducing their
contributions to a linear combination of \ tabulated Master integrals. This
obstacle only allowed us to calculate their leading term in the expansion in
$\log (\frac m\mu ).$ \ However, for $D_{32}$ it was possible to express it
as a sum over the Master integrals \ given in Ref. \cite{schavuor}.\ The
analytic expression of the diagram is
\begin{eqnarray}
\Gamma ^{(32)} &=&-V^{(d)}\frac 14(g_Y)^4\int \frac{dp_1^d}{(2\pi )^di}\frac{%
dp_2^d}{(2\pi )^di}\frac{dp_3^d}{(2\pi )^di}\times   \nonumber \\
&&\frac{\text{Tr}\left[ (m+p_2^\mu \gamma _\mu )(m+(p_2^\mu +p_3^\mu
-p_1^\mu
)\gamma _\mu )(m+p_3^\mu \gamma _\mu )(m+p_1^\mu \gamma _\mu )\right] }{%
(m^2-p_1^2)(m^2-p_2^2)(m^2-p_3^2)(m^2-(p_2+p_3-p_1)^2)(p_1-p_3)^2(p_1-p_2)^2}
\nonumber \\
&=&-V^{(d)}\frac 14(g_Y)^4\int \frac{dp_1^d}{(2\pi )^di}\frac{dp_2^d}{(2\pi
)^di}\frac{dp_3^d}{(2\pi )^di}\times   \nonumber \\
&&\frac{m^4+c_1(p_1,p_2,p_3)m^2+c_2(p_1,p_2,p_3)}{%
(m^2-p_1^2)(m^2-p_2^2)(m^2-p_3^2)(m^2-(p_2+p_3-p_1)^2)(p_1-p_3)^2(p_1-p_2)^2}%
, \\
c_1(p_1,p_2,p_3) &=&3p_2.p_{3+}p_1.p_2+p_1.p_3+p_2^2+p_3^2-p_1^2\text{ \ \ \
\ } \\
c_2(p_1,p_2,p_3) &=&p_1^2\text{ }p_2.p_3+p_2^2\text{ }p_1.p_3+p_3^2\text{ }%
p_1.p_2-2\text{ }p_1.p_2\text{ }p_1.p_3.
\end{eqnarray}

Defining now
\begin{eqnarray}
z_1 &=&p_1^2-m^2,  \nonumber \\
z_2 &=&p_2^2-m^2,  \nonumber \\
z_3 &=&p_3^2-m^2,  \nonumber \\
z_4 &=&(p_1-p_2)^2,  \nonumber \\
z_5 &=&(p_1-p_3)^2,  \nonumber \\
z_6 &=&(p_2-p_1+p_3)^2-m^2,
\end{eqnarray}
and employing various vectorial identities expressing the squares of the
differences between any two momenta in terms of the scalar product between
them and the squares of the considered momenta, the integral defining $%
\Gamma ^{(32)}$ can be written as follows
\begin{eqnarray}
\Gamma ^{(32)} &=&-V^{(d)}\frac 14(g_{Y\text{ }})^4\int \frac{dp_1^d}{(2\pi
)^di}\frac{dp_2^d}{(2\pi )^di}\frac{dp_3^d}{(2\pi )^di}\times   \nonumber \\
&&\frac{m^4+c_1(z)\text{ }m^2+c_2(z)}{z_1\text{ }z_2\text{ }z_3\text{ }z_4%
\text{ }z_5\text{ }z_6},  \nonumber \\
z &=&(z_1\text{, }z_2,z_3,\text{ }z_4,\text{ }z_5\text{,}z_6), \\
c_1(z) &=&\frac 32(z_1+\text{ }z_2+\text{ }z_3+z_6)-2(\text{ }z_4\text{ }%
+z_5)\text{ }+6m^2, \\
c_2(z) &=&\frac 12(z_1z_6+z_2z_3-z_4z_5+m^2(z_1+z_2+z_3+z_6)+2m^4).
\end{eqnarray}

Therefore, there are one or two $z$ factors in the denominator that can be
canceled by the terms of the quadratic polynomial in these quantities. This
allows the integral to be decomposed in a linear combination of the Master
integrals listed in Ref. \cite{schavuor}. The result for $\ $\ action density
\[
\gamma _{32}(m,\mu ,\epsilon )=\frac{\Gamma ^{(32)}}{\mu ^{2\epsilon }V^{(d)}%
},
\]
can be expressed in terms of only five of them as follows
\[
\gamma _{32}(m,\mu ,\epsilon )=-(g_{Y\text{ }}^0)^4\text{ }m^4\left( \frac
m\mu \right) ^{-6\epsilon }\left( 8I_1(\epsilon )+8I_2(\epsilon
)-4I_3(\epsilon )+I_5(\epsilon )-\frac{I_7(\epsilon )}2\right) ,
\]
where the functions $I_1(\epsilon )$, $I_2(\epsilon ),I_3(\epsilon
),I_5(\epsilon )$ and $I_7(\epsilon )$ result to be given by

\begin{eqnarray}
I_1(\epsilon )&=&\frac{2^{-3(4-2\epsilon )-9}\pi ^{-\frac 32(4-2\epsilon
)}(5(4-2\epsilon )-18)M_1(\epsilon )^3}{1-2\epsilon }+  \nonumber \\
&&\frac{2^{-3(4-2\epsilon )-6}\pi ^{-3(4-2\epsilon )}(3(4-2\epsilon
)-10)(3(4-2\epsilon )-8)\left( M_5(\epsilon )-\frac{8\epsilon }{%
2(4-2\epsilon )-7}M_4(\epsilon )\right) }{\epsilon ^2}, \\
I_2(\epsilon ) &=&-\frac{2^{-3(4-2\epsilon )-2}\pi ^{-3(4-2\epsilon )}}{%
1-2\epsilon }\left( \frac{M_1(\epsilon )^3(2-2\epsilon )^2}{1-2\epsilon }%
+(3(4-2\epsilon )-8)M_4(\epsilon )\right) , \\
I_3(\epsilon ) &=&-\frac{2^{-3(4-2\epsilon )-3}\pi ^{-3(4-2\epsilon )}}%
\epsilon \left( \frac{2(2-2\epsilon )^2M_1(\epsilon )^3}{1-2\epsilon }%
+(3(4-2\epsilon )-8)M_5(\epsilon )\right) , \\
I_5(\epsilon ) &=&(2\pi )^{-3(4-2\epsilon )}M_4(\epsilon ), \\
I_7(\epsilon ) &=&(2\pi )^{-3(4-2\epsilon )}M_5(\epsilon ),
\end{eqnarray}

in terms of the Master integrals (See Ref. \cite{schavuor}):

\begin{eqnarray}
M_1(\epsilon ) &=&\pi ^{\frac 12(4-2\epsilon )}\Gamma \left( \frac
12(2\epsilon -4)+1\right) , \\
M_2(\epsilon ) &=&-\frac{(2-2\epsilon )\text{ }M_1(\epsilon )^2}{%
2(1-2\epsilon )}, \\
M_3(\epsilon ) &=&2^{\frac 12(2\epsilon -4)}\Gamma \left( \frac
12(4-2\epsilon )\right) \Gamma \left( \frac 12(2\epsilon -1)\right)
M_1(\epsilon )^2, \\
M_4(\epsilon ) &=&\frac{2^{1-2\epsilon }\Gamma \left( \frac
12(8-3(4-2\epsilon ))\right) \Gamma \left( \frac 12(2\epsilon -1)\right) }{%
\Gamma \left( \frac 12(7-2(4-2\epsilon ))\right) \Gamma \left( \frac
12(2\epsilon -2)\right) }M_1(\epsilon )^3, \\
M_5(\epsilon ) &=&(-2-\frac 53\epsilon \text{ \ \ }-\text{\ }\frac
12\epsilon ^2\text{ \ +\ \ \ }\frac{103}{12}\epsilon ^3\text{ \ \ \ +\ \ }%
\frac 7{24}(163-128\zeta (3))\epsilon ^4\text{ }+  \nonumber \\
&&\text{\ \ }(\ \frac{9055}{48}+\frac{136\pi ^4}{45}+\frac 13(\pi ^2-\log
(2)^2)(32\log (2)^2)-168\zeta (3)\text{ \ } \\
&&-256\text{Li}_4(\frac 12) \,\,\text{}) \, \epsilon ^5\text{ })\text{ }%
M_1(\epsilon )^3,  \nonumber
\end{eqnarray}
where the special functions Li$_n(\frac 12)$ and $\zeta (n)$ are defined as
\begin{eqnarray}
\text{Li}_n(x) &=&\sum_{k=1}^\infty \frac 1{2^kk^n}, \\
\zeta (n) &=&\sum_{k=1}^\infty \frac 1{k^n}.
\end{eqnarray}

\ Finally, the application of the before described MS  procedure \
leads to the following formula for the contribution to the vacuum
effective action density of the diagram $D_{32}$

\begin{eqnarray}
&&\gamma _{32}(m,\mu)=\text{ \ }(g_Y^0)^4m^4\text{ }{\Huge (}%
0.0000329114\log ^5\left( \frac m\mu \right) -0.000105904\log ^4\left( \frac
m\mu \right) +0.0000165851\log ^3\left( \frac m\mu \right) +  \nonumber \\
&&0.000441159\log ^2\left( \frac m\mu \right) -0.00074347\log \left( \frac
m\mu \right) +0.000388237{\Huge )}.
\end{eqnarray}

It can be noted that this term has a high quintic power of $\log
^5(\frac m\mu )$ \ which is also determined by the high pole of the
$\epsilon $ expansion present in the function $I_1$. This is the
highest power of the $\log \left( \frac m\mu \right) $ expansion \
appearing in the results. The next  higher power, the fourth one,
also is arising in this term.

\subsubsection{Diagram \ D$_{31}$}

\ We were not able to exactly evaluate this contribution (and also the one
associated to D$_{33})$ in terms of Master integrals. Therefore, \ for both
of these terms we here limited ourself to evaluate their leading terms in
the expansion in powers of $\log \left( \frac m\mu \right) $. For this
purpose, the use was made of the circumstance that (at variance with $D_{32}$%
, but in coincidence with $D_{33}$) this term corresponds to a loop
formed by two one loop self-energy insertions. Since these
self-energy terms are explicitly calculable in terms of
hypergeometric functions, both terms can be expressed as single
momentum integral in $d$ dimensions. The diagram has the original
analytic expression
\begin{eqnarray}
\Gamma ^{(31)} &=&-V^{(d)}\frac 12(g_Y)^4\int \frac{dp_1^d}{(2\pi )^di}\frac{%
dp_2^d}{(2\pi )^di}\frac{dp_3^d}{(2\pi )^di}\times   \nonumber \\
&&\frac{Tr\left[ (m+p_2^\mu \gamma _\mu )(m+p_1^\mu \gamma _\mu )(m+p_3^\mu
\gamma _\mu )(m+p_1^\mu \gamma _\mu )\right] }{%
(m^2-p_1^2)^2(m^2-p_2^2)(m^2-p_3^2)(p_1-p_3)^2(p_1-p_2)^2}  \nonumber \\
&=&-V^{(d)}\frac 12(g_Y)^4\int \frac{dp_1^d}{(2\pi )^di}\frac{dp_2^d}{(2\pi
)^di}\frac{dp_3^d}{(2\pi )^di}\times   \nonumber \\
&&\frac{m^4+d_1(p_1,p_2,p_3)m^2+d_2(p_1,p_2,p_3)}{%
(m^2-p_1^2)^2(m^2-p_2^2)(m^2-p_3^2)(p_1-p_3)^2(p_1-p_2)^2}, \\
d_1(p_1,p_2,p_3) &=&p_1^2+2p_1.p_2+2p_1.p_3+p_2.p_3\text{ \ \ ,\ \ } \\
d_2(p_1,p_2,p_3) &=&2\text{ }p_1.p_2\text{ }p_1.p_3-\text{\ }p_1^2\text{ }%
p_2.p_3.
\end{eqnarray}

We define now the fermion self-energy integral and its related
vector as follows

\begin{eqnarray}
s_{31}(p^2) &=&\int \frac{dp_1^d}{(2\pi )^di}\frac 1{(m^2-p_1^2)(p_1-p)^2}
\nonumber \\
&=&-\frac{\pi ^{\frac d2}}{(2\pi )^d}\Gamma (\epsilon )\int_0^1dx\text{ }%
x^{-\epsilon }(m^2-p^2(1-x)-i\delta )^{-\epsilon }  \nonumber \\
&=&\frac{\pi ^{\frac d2}}{(2\pi )^d}\Gamma (\epsilon )(m^2)^{-\epsilon }%
\frac{_2F_1(1-\epsilon ,\epsilon ,2-\epsilon ,-\frac{(\frac pm)^2}{1-(\frac
pm)^2})}{\epsilon -1}, \\
v_\mu (p^2) &=&\int \frac{dp_1^d}{(2\pi )^di}\frac{p_{1\mu }}{%
(m^2-p_1^2)(p_1-p)^2}  \nonumber \\
&=&a(p^2)\text{ \ }p_\mu , \\
a(p^2) &=&\frac{p^2+m^2}{2p^2}s_{31}(p^2)-\frac{L(m,\epsilon )}{2p^2}, \\
L(m,\epsilon ) &=&\int \frac{dp_1^d}{(2\pi )^di}\frac 1{(m^2-p_1^2)}=\frac{%
\pi ^{\frac d2}m^{d-2}}{(2\pi )^d}\Gamma (1-\frac d2).
\end{eqnarray}

In the above expressions, the Feynman \ parametric integral was
explicitly evaluated by employing the algebraic calculation program
Mathematica. After performing the Wick rotation in the external and
integration momenta  and extracting the $d$-dimensional solid angle
arising form the angular integrals, the expression for the action
density can be written as follows
\[
\ \gamma _{31}(m,\mu ,\epsilon )=\frac{\Gamma ^{(31)}}{\mu ^{2\epsilon
}V^{(d)}},
\]
get the expression \ \
\begin{eqnarray}
\gamma _{31}(m,\mu ,\epsilon ) &=&-\frac{2(g_Y^0)^4\left( \frac m\mu \right)
^{-6\text{ }\epsilon }c(m,\epsilon )}{\epsilon ^2}\int_0^\infty \frac{%
p^{3-2\epsilon }}{\left( p^2+1\right) ^2}\,f(p,\epsilon )dp, \\
f(p,\epsilon ) &=&\epsilon ^2f_1(p,\epsilon )\Gamma (\epsilon
)^2+f_2(p,\epsilon )\text{ }\epsilon \text{ }\Gamma (\epsilon
)+f_3(p,\epsilon ), \\
f_1(p,\epsilon ) &=&(1-p^2)(3-\frac{(1-p^2)^2}{4p^2})(s_{31}^{*}(p^2,%
\epsilon ))^2, \\
f_2(p,\epsilon ) &=&(2-\frac{(1-p^2)^2}{2p^2})s_{31}^{*}(p^2,\epsilon
)L^{*}(\epsilon ), \\
f_3(p,\epsilon ) &=&-\frac{\left( 1-p^2\right) ^2(L^{*}(\epsilon ))^2}{4p^2},
\\
s_{31}^{*}(p,\epsilon ) &=&-\frac{2^{2\epsilon -4}\pi ^{\frac 12(4-2\epsilon
)+2\epsilon -4}\,_2F_1\left( 1-\epsilon ,\epsilon ;2-\epsilon ;\frac{p^2}{%
p^2+1}\right) }{\epsilon -1}, \\
c(m,\epsilon ) &=&\frac{2^{2\epsilon -3}m^4\pi ^{\frac 12(4-2\epsilon
)+2\epsilon -4}}{\Gamma \left( \frac 12(4-2\epsilon )\right) }, \\
L^{*}(\epsilon ) &=&\epsilon L(1,\epsilon ).
\end{eqnarray}

\ \ As it was mentioned before, we were not able yet to find an epsilon
expansion (rigorous or sufficiently approximated numerical one) allowing to
exactly evaluate this integral after removing the regularization. Therefore,
in order to determine an approximation for $\gamma _{31}$ we have made use
of an assumption suggested by an exploration done about the asymptotic power
expansion at infinity of the integrand as a function of the momentum
integration variable $p$. \ It followed that all the terms of the expansion
\ after integrated, show a single pole structure in their Laurent expansion
in $\epsilon .$ Then, it suggests that the full \ divergence of the integral
at $d=4$ is defined by a single pole in $\epsilon .$ \ Assuming this
property, the extraction of the leading correction in $\log (\frac m\mu )$
should be defined by the maximal power of $\log (\frac m\mu )$ \ appearing
in the coefficient of the zero order term in the expansion of the modified
integral
\begin{equation}
\gamma _{31}(m,\mu ,\epsilon )=-\frac{2(g_Y^0)^4\left( \frac m\mu \right)
^{-6\text{ }\epsilon }c(m,0)}{\epsilon ^2}\int_0^\infty \frac{p^{3-2\epsilon
}}{\left( p^2+1\right) ^2}\,f(p,0)dp.
\end{equation}

Note that any other power of $\epsilon $ in the expansions of $c(m,\epsilon
) $ and $\ f(p,\epsilon )$ \ will reduce the maximal order of the negative
powers of epsilon in the full expansion of $\gamma _{31}(m,\mu ,\epsilon )$
which determines the leading correction in the expansion. \ \ For $\,f(p,0)$
\ it follows
\begin{equation}
\,f(p,0)=\frac{p^4}{1024\pi ^4}-\frac{17p^2}{1024\pi ^4}+\frac 7{256\pi
^4}-\frac 1{256\pi ^4p^2}.
\end{equation}

Then, the use of the formula
\begin{equation}
\int_0^\infty \frac{p^{3-2\epsilon+m }}{\left( p^2+1\right) ^2}dp=-\frac \pi
4(m-2\epsilon +2)\csc (\frac \pi 2(m-2\epsilon )),
\end{equation}
which shows the $\frac 1\epsilon $ singularity, allows to write for $\gamma
_{31}$ the leading logarithm correction to its finite part
\begin{equation}
\gamma _{31}(m,\mu )=-0.0000228551(g_Y^0)^4\text{ }m^4\log ^3\left( \frac
m\mu \right) .
\end{equation}

\subsubsection{ Diagram $D_{33}$}

\ \ As it was remarked, this terms will be treated in a similar way
as it was done for $D_{31}$. Now, the corresponding self-energy
insertions will be the boson ones. Again, the two self-energy loops
are explicitly calculable in terms of hypergeometric functions. The
starting analytic expression of the diagram is \
\begin{eqnarray}
\Gamma ^{(33)} &=&\frac 14V^{(d)}(g_Y)^4\int \frac{dp^d}{(2\pi )^di}\frac{%
dp_1^d}{(2\pi )^di}\frac{dp_2^d}{(2\pi )^di}\times \frac 1{(p^2)^2}
\nonumber \\
&&\frac{\text{Tr}[(m+p_1^\mu \gamma _\mu )(m+(p+p_1)^\nu \gamma _\nu
)]\text{Tr}[(m+p_2^\mu \gamma _\mu )(m+(p_2+p)^\nu \gamma _\nu )]}{%
(m^2-p_1^2)^2(m^2-(p_1+p)^2)(m^2-p_2^2)^2(m^2-(p_2+p)^2)},  \nonumber \\
&=&4V^{(d)}(g_Y)^4\int \frac{dp^d}{(2\pi )^di}\frac{dp_1^d}{(2\pi )^di}\frac{%
dp_2^d}{(2\pi )^di}\times \frac 1{(p^2)^2}\times   \nonumber \\
&&\frac{(m^2+p_1.(p_1+p))(m^2+p_2.(p_2+p))}{%
(m^2-p_1^2)^2(m^2-(p_1+p)^2)(m^2-p_2^2)^2(m^2-(p_2+p)^2)},
\end{eqnarray}
where the fermion traces were evaluated for writing the second form
of the integral. The last expression evidences the decomposition in
two serial self-energy terms. \

After rotating to Euclidean space the momenta variables of the integration
regions and the external momentum, the fermion selfenergy integral and its
related vector integral can be written as follows (See Ref. \ \cite{muta})
\begin{eqnarray*}
s_{33}(q^2,\epsilon ) &=&\int \frac{dq_1^d}{(2\pi )^d}\frac
1{(m^2+q_1^2)(m^2+(q+q_1)^2} \\
&=&\frac{(m)^{-2\epsilon }}{(4\pi )^{\frac d2}}\Gamma (\epsilon )\int_0^1dx%
\text{ }(1+(\frac qm)^2x(1-x))^{-\epsilon } \\
&=&\frac{(m)^{-2\epsilon }}{(4\pi )^{\frac d2}}\Gamma (\epsilon )\mathcal{F(}%
\frac{q^2}{m^2}), \\
\mathcal{F(}q^2) &=&\int_0^1dx\text{ }(1+q^2x(1-x))^{-\epsilon } \\
&=&-\frac{2^{-\epsilon -1}(q+\sqrt{q^2+4})(1-\frac q{\sqrt{q^2+4}})^\epsilon \,
_2F_1(1-\epsilon ,\epsilon ,2-\epsilon ,\frac 12(\frac q{\sqrt{q^2+4}}+1))}{%
q(\epsilon -1)}- \\
&&\frac{2^{-\epsilon -1}(q-\sqrt{q^2+4})(1+\frac q{\sqrt{q^2+4}})^\epsilon
\, _2F_1(1-\epsilon ,\epsilon ,2-\epsilon ,\frac 12(-\frac q{\sqrt{q^2+4}}+1))}{%
q(\epsilon -1)}, \\
v_{33\mu }(p^2) &=&\int \frac{dp_1^d}{(2\pi )^di}\frac{p_{1\mu }}{%
(m^2-p_1^2)(m^2-(p+p_1)^2} \\
&=&a(p^2)\text{ \ }p_\mu , \\
a(p^2) &=&-\frac 12s_{33}(p^2,\epsilon ).
\end{eqnarray*}

Again the result for parametric Feynman integral was analytically evaluated
thanks to the use of the algebraic calculation program Mathematica.

Thus, after extracting the Euclidean angular integrals and performing some
transformations, the density of the effective action
\[
\ \gamma _{33}(m,\mu ,\epsilon )=\frac{\Gamma ^{(33)}}{\mu ^{2\epsilon
}V^{(d)}},
\]
can be expressed as a single momentum integral in the range (0,$\infty $) as
follows
\begin{eqnarray}
\gamma _{33}(m,\mu ,\epsilon ) &=&\frac{4c(m,\epsilon )(g_Y^0)^4\left( \frac
m\mu \right) ^{-6\epsilon }}{\epsilon ^2}\int_0^\infty dp\text{ }\frac{%
p^{3-2\epsilon }}{\left( p^2+r^2\right) ^2}g(p,\epsilon ), \\
g(p,\epsilon ) &=&\epsilon ^2g_1(p,\epsilon )\Gamma (\epsilon
)^2+g_2(p,\epsilon )\text{ }\epsilon \text{ }\Gamma (\epsilon
)+g_3(p,\epsilon ), \\
g_1(m,\epsilon ) &=&\left( \frac{p^2}2+2\right) ^2s_{33}^{*}(p,\epsilon )^2,
\\
g_2(m,\epsilon ) &=&-2\left( \frac{p^2}2+2\right) L(\epsilon )\text{ }%
s_{33}^{*}(p,\epsilon ), \\
g_3(m,\epsilon ) &=&L(\epsilon )^2, \\
s_{33}^{*}(p,\epsilon ) &=&\frac{s_{33}(p,\epsilon )}{\Gamma (\epsilon )}, \\
c(m,\epsilon ) &=&\frac{2^{2\epsilon -3}m^4\pi ^{\frac
12(4-2\epsilon )+2\epsilon -4}}{\Gamma \left( \frac 12(4-2\epsilon
)\right) }, \\
L(\epsilon ) &=&\frac{%
\pi ^{2-\epsilon}}{(2\pi )^{4-2\epsilon}}\Gamma (-1+\epsilon).
\end{eqnarray}

Finally, by employing a similar procedure for extracting the leading
logarithmic correction in $\log \left( \frac m\mu \right) $ for $D_{31}$,
the analogous contribution for $D_{33}$ follows in the form \
\[
\gamma _{33}(m,\mu )=-0.000329114\text{ }(g_Y^0)^4\text{ }m^4\log
^3\left( \frac m\mu \right).
\]

\section{Discussion}

\ Lets us now comment the \ results obtained in previous sections for the
effective action density. \ The total effective potential \ value $v(m,\mu
), $ is given by the sum of all the evaluated terms after changing their
sign. The total potential and its various contributions are written below
\begin{eqnarray}
v(m,\mu ) &=&v_1(m,\mu )+v_2(m,\mu )+v_{31}(m,\mu )+v_{33}(m,\mu
)+v_{32}(m,\mu ), \\
\frac{v_1(m,\mu )}{m^4} &=&-\frac{\gamma _1(m,\mu )}{m^4}=-0.0506606%
\left( 2.\log \left( \frac m\mu \right) -2.95381\right) , \\
\frac{v_2(m,\mu )}{m^4} &=&-\frac{\gamma _2(m,\mu )}{m^4}%
=-0.0000200507(g_Y^0)^2(183.83-173.783\log \left( \frac m\mu \right)
+48.\log ^2\left( \frac m\mu \right) ), \\
\frac{v_{31}(m,\mu )}{m^4} &=&-\frac{\gamma _{31}(m,\mu )}{m^4}%
=0.0000228551(g_Y^0)^4m^4\log ^3\left( \frac m\mu \right) , \\
\frac{v_{33}(m,\mu )}{m^4} &=&-\frac{\gamma _{33}(m,\mu )}{m^4}%
=0.000329114(g_Y^0)^4m^4\log ^3\left( \frac m\mu \right) , \\
\frac{v_{32}(m,\mu )}{m^4} &=&-\frac{\gamma _{32}(m,\mu )}{m^4}=\text{ \ }%
-(g_Y^0)^4m^4\text{ }10^{-3}{\Huge (}0.0329114\log ^5\left( \frac m\mu
\right) -0.105904\log ^4\left( \frac m\mu \right) +  \nonumber \\
&&0.0165851\log ^3\left( \frac m\mu \right) +0.441159\log ^2\left( \frac
m\mu \right) -0.74347\log \left( \frac m\mu \right) +0.388237{\Huge )}.
\end{eqnarray}

Let us now \ consider that the renormalization point for $\mu $ is
chosen at the same value of the fermion mass $m_f,$ under
consideration, that is $\ Log(\frac{m_f}\mu )=0$. Also we will
define new scaled scalar field $\Phi $ and interaction parameter $g$
by mean of
\begin{eqnarray}
\Phi &=&\alpha \varphi , \\
g_Y^0 &=&\alpha m=g\exp (\Phi ), \\
g &=&\alpha m_f,
\end{eqnarray}
Then, the evaluated total contribution to the effective potential for the
Dilaton $v(m,\mu )$ can be expressed as a function $v(\Phi ,g)$ as follows
\begin{eqnarray}
\frac{v(\Phi ,g)}{m_f^4} &\equiv &\frac{v(m,\mu )}{m_f^4}=\text{ \ }%
-0.0000329114e^{8\Phi }g^4\Phi ^5+0.000105904e^{8\Phi }g^4\Phi ^4  \nonumber
\\
&&+0.000289673 e^{8\Phi }g^4\Phi ^3+e^{4\Phi }\left(
-0.000441159e^{4\Phi
}g^4-0.000962436e^{2\Phi }g^2\right) \Phi ^2+  \nonumber \\
&&e^{4\Phi }\left( 0.00074347e^{4\Phi }g^4+0.00348448e^{2\Phi
}g^2-0.101321\right) \Phi +  \nonumber \\
&&e^{4\Phi }\left( -0.000388237e^{4\Phi }g^4-0.00368594e^{2\Phi
}g^2+0.149642\right) .
\end{eqnarray}

\begin{figure}[tbp]
\includegraphics[width=7.5cm]{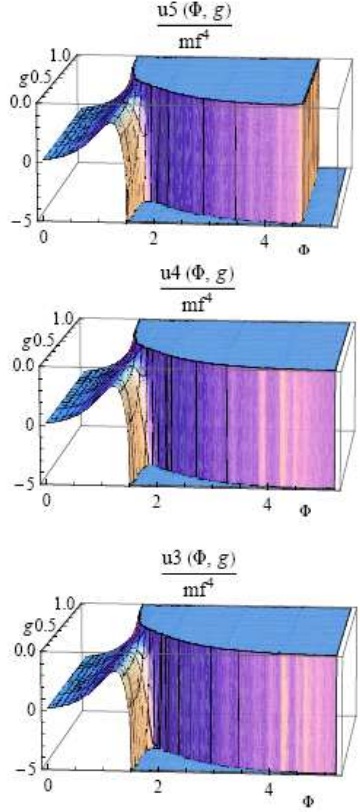}
\caption{ The three figures show, from top to bottom, the potentials $u_5$, $%
u_4$ and $u_3$ dependence on the field $\Phi $ and the coupling $g$,
respectively. The potential scale is chosen for a high magnification
range (the minima of the surface at fixed $g$ values are very far
below the plotted range) in order to evidence the presence of a
threshold for the appearing of the minima when the value of $g$
decreases below $g=1$. Note that for $\Phi $ smaller than some units
and not to small values of $g$, the three plotted graphs are
similar, indicating that the elimination of the highest fifth, and
also the next to highest fourth, powers of the field (or, of the
logarithm in the original expansion) in defining $u_4$ and $u_3$
respectively, are not affecting the results in the mentioned region.
The circumstance that the exact evaluated contribution has a
negative leading term of order five (which makes the result
unbounded from below) explains that for the plot of $u_5$ the minima
disappear for sufficiently small values of $g$. However, the fact
that exact potential should be expected to be bounded from below, we
consider that supports our assumption about employing the bounded
from below approximations of the potential $u_4$  in evaluating the
Dilaton properties at the small values of $g$ defined by the $GUT$
and $m_{top}$ mass scales.} \label{figura2}
\end{figure}

Let us also define now the functions $u_{5,}$ $u_4$ and $u_3$ in the
following form
\begin{eqnarray}
\frac{u_5(\Phi ,g)}{m_f^4} &=&\frac{v(\Phi ,g)}{m_f^4}, \\
\frac{u_4(\Phi ,g)}{m_f^4} &=&\text{ \ }0.000105904e^{8\Phi }g^4\Phi
^4+0.000289673 e^{8\Phi }g^4\Phi ^3+  \nonumber \\
&&e^{4\Phi }\left( -0.000441159e^{4\Phi }g^4-0.000962436e^{2\Phi
}g^2\right)
\Phi ^2+  \nonumber \\
&&e^{4\Phi }\left( 0.00074347e^{4\Phi }g^4+0.00348448e^{2\Phi
}g^2-0.101321\right) \Phi +  \nonumber \\
&&e^{4\Phi }\left( -0.000388237e^{4\Phi }g^4-0.00368594e^{2\Phi
}g^2+0.149642\right) ,  \label{u4} \\
\frac{u_3(\Phi ,g)}{m_f^4} &=&\text{ \ }+0.000289673 e^{8\Phi
}g^4\Phi ^3+
\nonumber \\
&&e^{4\Phi }\left( -0.000441159e^{4\Phi }g^4-0.000962436e^{2\Phi }g^2\right)
\Phi ^2+  \nonumber \\
&&e^{4\Phi }\left( 0.00074347e^{4\Phi }g^4+0.00348448e^{2\Phi
}g^2-0.101321\right) \Phi +  \nonumber   \\
&&e^{4\Phi }\left(
-0.000388237e^{4\Phi }g^4-0.00368594e^{2\Phi }g^2+0.149642\right) .
\end{eqnarray}
Note that $u_5$ coincides $v$ and is of order five in the powers of $\Phi $.
The function $u_4$, $u_3$ are defined as retaining only all the terms up to
order $\Phi ^4$ and $\Phi ^3$ respectively of the original function $u_5$ .
\ Therefore, these functions basically correspond to the expansion of order
five, four and three in powers of $\log \left( \frac m\mu \right) .$ \ They
are defined in order to study the influence of increasing the order of the
perturbative expansion in powers of $\ \log \left( \frac m\mu \right) .$

\ To evidence the dependence on $\Phi $ and $g$ of the three functions
(after divided by the common factor $m_f^4$), they are plotted in figure \ref
{figura2}\ . \ The range of values of $g=m_f\alpha $ \ was chosen $(0,1)$\
as suggested by the fact that $\alpha $ is of the order of the Planck length
and thus the physical values of the considered fermion mass are expected to
determine $g$ to be smaller than one. \ The plot of $u_5$ shows that there
is a threshold value of $g$, \ below which the \ potential shows minima
tending to stabilize the \ vacuum mean value of the Dilaton field. \ This
behavior is also shown by the approximated potentials $u_4$ and $u_3$, a
fact that indicates that after disregarding the higher quintic and quartic
terms in the expansion in $Log(\frac m\mu $), the existence of Dilaton
stabilizing minima is not affected. \ \

When considering the full evaluated potential curve $u_5,$
illustrated at the top plot of figure \ref{figura2},\ it can be
observed \ that after lowering the $g$ value below \ a critical
threshold, the minimum as a function of $\Phi $ stops to exist at a
critical value $g_{min}.$ However,
in the case of $u_4$ and $u_3$ the minimum exists for arbitrary values of $%
g->0$. That is, when the potential approximations is bounded from
below, the potential shows stabilizing minima at any small value of
$g$ close to zero. The field value at the minima grow when the
coupling tends to vanish.\ It can be noted, that the non bounded
from below \ character of the approximated potential calculated here
is determined by the fact that the quintic power of $\Phi $
correction turns to be negative. However, the physical system under
consideration is one in which the total effective potential can be
expected to show an exact bounded from below character. Thus, the
next corrections are expected to exhibit a bounded from below
behavior. In accordance with this expectation, in studying the $g$
dependence at small values, we will employ the bounded from below
approximated potential function $u_4$, assuming that it represents a
reasonably good approximation of the exact potential.

\subsection{Dilaton field and mass for $m_f$ at the $GUT$ scale}

\begin{figure}[tbp]
\begin{center}
\hspace*{-0.4cm} \includegraphics[width=7.5cm]{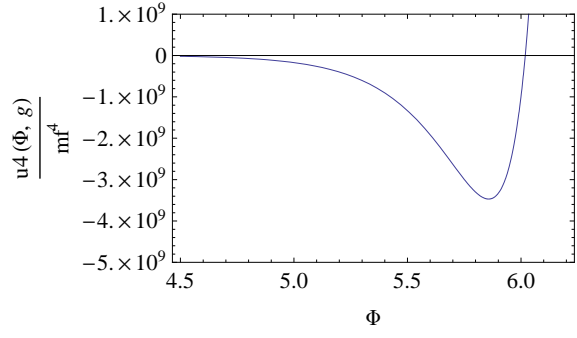}
\end{center}
\caption{ The effective potential $u_4$ defined by Eq. \ref{u4} as a
function of the Dilaton field $\Phi $. The fermion mass was fixed to
correspond to the $GUT$ mass $m_{GUT}$ and the renormalization scale $\mu $
was chosen to coincide with this mass. The minimum of the potential is near
the value $\Phi =5.7$, which indicates that the field is bound to a high
value near the Planck scale. }
\label{figura3}
\end{figure}

Let us consider now that the highest fermion mass $m_f$ is given by the $GUT
$ mass scale

\begin{eqnarray}
m_f &=&m_{GUT}=5.06773\times 10^{29}\text{cm}^{-1}  \nonumber \\
&\equiv &10^{16}\ \text{GeV},
\end{eqnarray}
which produces for the coupling $g$ the \ value
\[
g=m_f\text{ }\alpha =-\frac 34\kappa \text{ }m_{GUT}=-0.0030789542773.
\]
\ The potential $u_4$ as function of the field $\Phi $ for this particular
value of $g$ is shown in \ figure \ \ref{figura3}. \ \ The minimum of the
curve determines an estimate for the vacuum value of the Dilaton field given
by \ \
\begin{eqnarray}
\Phi _{vac} &=&5.8576156\text{ }= \alpha \text{ }\varphi _{vac}, \\
\text{ }\varphi _{vac} &=&-\frac 43 {5.8576156}\frac 1\kappa .
\end{eqnarray}
This result indicates that the \ vacuum mean value of the Dilaton field,
after assuming that the fermion mass is in the $GUT$ scale, becomes
stabilized in the scale of the Planck mass. \

Let us consider now the mass of the field excitation. \ Its value is
determined by the second derivative of the potential curve taken at the
minimum, which is given by
\begin{equation}
\left. \frac{d^2}{d\Phi ^2}u_4(\Phi ,-0.0030789542773)\right| _{\Phi
=\Phi^{(m_{GUT})} _{vac}}=1.28179\times 10^{11}m_f^4.
\end{equation}

In order to \ estimate the Dilaton mass let us consider the \ linearized \
equation of motion for the mean field
\begin{equation}
(\frac 1{\alpha ^2}\partial ^2+\left. \frac{d^2}{d\Phi ^2}u_4(\Phi
,-0.00307895)\right| _{\Phi =\Phi _{vac}^{(m_{GUT})}})\, \Phi =0,
\end{equation}
in which the factor $\frac 1{\alpha ^2}$ multiplying the \ \ D'Alembertian \
appears due to the previously done change of field variable $\Phi =\alpha \,
\varphi.$

The above wave equation leads to the dispersion relation for the Dilaton
modes

\begin{equation}
(-\frac 1{\alpha ^2}p^2+\left. \frac{d^2}{d\Phi ^2}u_4(\Phi
,-0.00307895)\right| _{\Phi =\Phi _{vac}^{(m_{GUT})}})=0,
\end{equation}
which for the case of the particle at rest $p=(m_D^{(m_{GUT})},0,0,0)$ \
determines for the Dilaton the mass estimate
\begin{eqnarray}
m_D^{(m_{GUT})} &=&\sqrt{\left. \frac{d^2}{d\Phi ^2}u_4(\Phi
,-0.00307895)\right| _{\Phi =\Phi _{vac}^{(m_{GUT})}}}m_{GUT}^2\mid \alpha
\mid  \nonumber \\
&=&\text{ \ \ }5.58626\times 10^{32}\ \ \ \ \text{\ \ \ cm}^{-1}.
\end{eqnarray}

Therefore, the predicted order of the mass \ for the Dilaton also lays at
extremely high values \ which \ make this field mode undetectable in a
direct way. \

\subsection{Dilaton mean value and mass for $m_f$ at the $top$ quark mass
scale}

\begin{figure}[tbp]
\begin{center}
\hspace*{-0.4cm} \includegraphics[width=7.5cm]{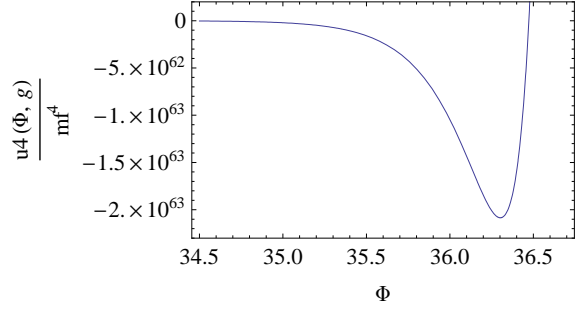}
\end{center}
\caption{ The effective potential $u_4$ plotted as a function of the Dilaton
field $\Phi $. In this case the coupling was defined by a fermion mass
correspond to the top quark one $m_{top}$ and the renormalization scale $\mu
$ was also chosen to coincide with this value. The minimum of the potential
is now near the value $\Phi =36.7765$, which indicates that the field is
again staying at a high value. }
\label{figura4}
\end{figure}

It is also of interest to take as $m_f$ the highest currently known fermion
mass: that is, the top quark one $\ $

\begin{equation}
m_{top}=172.0\pm 0.9\ \ \ \text{GeV}=8.7164\times 10^{15}\,\ \ \text{cm}%
^{-1}.
\end{equation}
Then, the coupling $g$ in this case takes the small value
\begin{equation}
g=m_f \, \alpha =-\frac 34 \kappa \, m_{top}=-5.32659 \times
10^{-17}.
\end{equation}

\ Figure \ref{figura4} shows the dependence of the potential $u_4$ as a
function of the field $\Phi $ at the above value of the coupling $g$. \ \
The minimum of the curve in this case gives for the mean Dilaton field \ at
the vacuum \ \
\begin{eqnarray}
\Phi _{vac}^{(m_{top})} &=&36.3020096=\alpha \text{ }\varphi
_{vac}^{(m_{top})}, \\
\text{ }\varphi _{vac}^{(m_{top})} &=&-\frac 43 {36.3020096}\frac
1\kappa .
\end{eqnarray}

This result predicts that, assuming that the maximal fermion mass in Nature
is given by the top quark one, which means a lower bound for the physical
masses, the vacuum field of \ the Dilaton, again becomes stabilized in a
scale, which although not being so high, is yet close to the Planck mass. \

\ In this case the dispersion relation for the Dilaton modes takes the form
\begin{equation}
(-\frac 1{\alpha ^2}p^2+\left. \frac{d^2}{d\Phi ^2}u_4(\Phi ,-5.32659\times
10^{-17})\right| _{\Phi =\Phi _{vac}^{(m_{top})}})=0.
\end{equation}

But, after evaluating for the second derivative of the potential at the
minimum to be \
\begin{equation}
\left. \frac{d^2}{d\Phi ^2}u_4(\Phi ,-5.32659\times 10^{-17})\right| _{\Phi
=\Phi _{vac}^{(m_{top})}}=6.86404\times 10^{64},
\end{equation}
and fixing again the rest frame momentum $p=(m_D^{(m_{top})},0,0,0)$ \
estimates for the Dilaton mass the value
\begin{eqnarray}
m_D^{(m_{top})} &=&\sqrt{\left. \frac{d^2}{d\Phi ^2}u_4(\Phi ,-5.32659\times
10^{-17})\right| _{\Phi =\Phi _{vac}^{(m_{top})}}}\times m_{top}^2\mid
\alpha \mid  \nonumber \\
&=&\text{ }\ 7.07209\times 10^{29}\text{\ \ \ \ \ \ \ \ \ \ \ \ \ \ \ cm}%
^{-1}.
\end{eqnarray}

Henceforth, also in this case the predicted mass for the Dilaton turns to be
a high value being now close to the $GUT$ scale. Thus, it can be expected
that for a maximal fermion mass in Nature ranging between the lower bound $%
m_{top} $ and the \ $GUT$ scale one, the \ \ Dilaton gets stabilized at a
large field value as required by string phenomenology. In addition the
resulting values of its mass, for the same range of $m_f$ masses, is also
out of the current observability range of particle detectors.

\section{Conclusions}

The predictions for the Dilaton stabilization problem determined the
existence of massive fermion matter had been further investigated here. The
fermion field mass values are considered in two cases: the top quark mass
representing the lower bound of all existing but yet unknown fermion masses
in Nature, and the energy scale of the grand unification theories of order $%
10^{16}$ GeV. In both situations, the results indicate that the Dilaton mean
field becomes stabilized at the very high values required by its role in
allowing gravity to have its observed properties. Then, the same existence
of matter seems to be a possible source of the dynamical fixation of the
Dilaton field at the high values, required by String Theory to imply the
observable Einstein theory of gravity. \ Furthermore, the evaluations
indicate that the Dilaton field is also found to be strongly bound around
its mean value, by showing a large mass being close to the $GUT$ or Planck
scales. Therefore, the work identifies a possible explanation for the lack
of observable consequences of the Dilaton scalar field in Nature. The
discussion included \ contributions to the effective potential up to
3-loops. This allows to consider the influence of the inclusion of different
leading perturbative correction on the main conclusions. After, disregarding
in the evaluated potential: a) the highest order term (quintic) in the
expansion in powers of $Log(\frac m\mu )$ (which determined the unbounded
from below structure of the potential at large $\Phi $ \ values ) \ or b)
the two highest orders (the quintic and the quartic ones), the obtained
modified potentials are both bounded from below at high field values. This
procedure allows that minima as functions of $\Phi $ exist for arbitrarily
small values of \ the coupling $g$. This allows to evaluate the small
coupling values associated to the $GUT$ and $top$ quark masses. The fact
that the Yukawa theory under consideration should exhibit a bounded from
below potential, then supports this here adopted procedure for \ estimating
the \ vacuum mean values and mass of the Dilaton field. However, further
higher loop evaluations are convenient to define more precise estimated
values of the Dilaton vacuum field and mass and also for checking that they
do not affect the picture.
  The validity of the employed Yukawa like approximation in which the power expansion in Dilaton
field of  exponential factor of the Fermion mass is limited to the linear
 term in the field, is argued when the fermion mass is very much smaller than Plank mass.

\begin{acknowledgments}
  One of the authors (A. C.) like to deeply acknowledge the
 support received from the Proyecto Nacional de Ciencias B\'asics
 (PNCB, CITMA, Cuba) and the Network N-35 of the Office of External
 Activities (OEA) of the ICTP (Italy). (M. R.) wish to acknowledge the
 support received from the Physics Department at the University of Helsinki,
(E. E.) thanks the support of the BCGS and the Physics Institute of Bonn University.
\end{acknowledgments}


\appendix
\section*{Appendix A}

\  In this Appendix we discuss  two points which improve the argues in the work:
a) Adding a discussion about the validity of the Yukawa approximation; and
b) To clarify the exposition of the results which indicate the stabilization of the vev and the mass generation.

Below  we address the  two  main  critical remarks we identified in the Report: \

\subsection{Yukawa approximation }

Let us study the
validity of only retaining the linear term in the expansion of the exponential of
the Dilaton field  multiplying the fermion mass in the action. For this purpose let us
consider the mean value of the square of the radiation field \ $\phi_{r}$
times the Dilaton coupling $\alpha$ in the just considered Yukawa
model. \ The square root of this quantity gives a measure of the amount of the
quantum fluctuations of $\alpha$ $\phi_{r}(x)$. A resulting small value of
this quantity justifies the approximation done in the work. In terms of the
Dilaton Green function $D$ this quantity can be written as follows:%
\begin{align}
\langle0|\text{ }\alpha_{(R)}^{2}\phi_{(R)}^{r2}(x)|0\rangle &  =\alpha
_{(R)}^{2}\int\frac{dp^{d}}{(2\pi)^{d}i}D(p)  \nonumber\\
&  =\alpha_{(R)}^{2}\int\frac{dp^{d}}{(2\pi)^{d}i}\frac{1}{-p^{2}+\Pi
(p,m,d)}\nonumber\\
&  =\alpha_{(R)}^{2}\int\frac{dp_{e}^{d}}{(2\pi)^{d}}\frac{1}{p_{e}^{2}%
+\Pi(p_{e},m,d)}.%
\end{align}
where $\Pi$ is its selfenergy which will be evaluated in the one loop
approximation, by excluding the divergent pole parts. This corresponds to
using the MS substraction by also employing the Feynman expansion in terms of
the renormalized fields incorporating counterterms.

\ Let us now consider that the bare Dilaton theory being quantized, is valid
in the Planck scale where its action is assumed to define a low energy
physics  for a string theory. Then, after denoting with the B subindex the
bare quantities, the relations between the bare and renormalized magnitudes are
\begin{align}
\alpha_{(B)}  &  =-\frac{3}{4}\kappa = Z_{\alpha}^{\frac{1}{2}}\alpha_{(R)}^{2}, \\
\phi_{(B)}^{r}(x)  &  =Z_{\phi}^{\frac{1}{2}}\phi_{(R)}^{r}(x).
\end{align}

\ We will also assume that the product $\alpha_{(R)}^{2}\phi_{(R)}^{r2}(x)$,
after the study of the renormalization properties of the theory, can
be chosen as a  renormalization group invariant, which determines for the
renormalization constants the relation
\[
Z_{\alpha}^{\frac{1}{2}}=\frac{1}{Z_{\phi}^{\frac{1}{2}}}.
\]

Therefore, the squared deviation of the field can be written in the form%
\begin{align}
\langle0|\alpha_{(R)}^{2}\phi_{(R)}^{r2}(x)|0\rangle &  =\frac{\alpha
_{(B)}^{2}}{Z_{\phi}}\langle0|\text{ }\phi_{(R)}^{r2}(x)\text{ }|0\rangle \nonumber\\
&  =\frac{\alpha_{(B)}^{2}}{Z_{\phi}}\int\frac{dp_{e}^{d}}{(2\pi)^{d}}\frac
{1}{p_{e}^{2}+\Pi^{(1)}(p_{e},m,d)}.%
\end{align}

Now, in order to get an estimate of this quantity, let us consider the
approximation in which the renormalized Yukawa coupling is small and also
assume that the finite part of the one loop selfenergy is evaluated at zero
momentum. As it will be seen in the comment of the next point, this quantity
determines the first approximation for the Dilaton mass as given by the square
root of the value of $\ p_{e}^{2}$ fixing the pole of the propagator. As it was mentioned,
in the discussion of the next point, assuming that the coupling is small implies that
the selfenergy evaluated at zero momentum approximately determines the pole mass
of the Dilaton. Thus in this small coupling limit at least
the above described approximation  can give a reasonable estimate of  the field
fluctuations. \ In addition, as it was supposed in the work,  we will fix the scale parameter $\mu$ as
coinciding with the fermion mass, that is $\mu=m.$ Therefore,
approximating $\Pi^{(1)}(p_{e},m,d)$ by $\Pi^{(1)}(0,m,d)$ and evaluating the
remaining simple momentum integral gives%
\begin{align}
\langle0|\alpha_{(R)}^{2}\phi_{(R)}^{r2}(x)|0\rangle &  =\frac{\alpha
_{(B)}^{2}}{Z_{\phi}}\int\frac{dp_{e}^{d}}{(2\pi)^{d}}\frac{1}{p_{e}^{2}%
+\Pi^{(1)}(0,m,d)} \nonumber\\
&  =\frac{\alpha_{(B)}^{2}}{Z_{\phi}}\text{ }\Pi^{(1)}(0,m,d)^{\frac{d-2}{2}%
}\frac{(\pi)^{2-\epsilon}\Gamma(\epsilon-1)}{(2\pi)^{4-2\epsilon}}.%
\label{Pi}
\end{align}

In order to proceed, let us evaluate the wavefunction renormalization constant
$Z_{\phi}$  to further transform the previous formula. The fermion one
loop contribution without the substractions is given by the expression
\[
\Pi(p,m,d)=-g_{y}^{2}\int\frac{dq^{d}}{(2\pi)^{D}i}Tr[\frac{1}%
{m-\gamma_{\mu}p^{\mu}}\frac{1}{m-\gamma_{\mu}(p-q)^{\mu}}],
\]
to which should be added the wavefunction and mass first order counterterms in
order to evaluate $\Pi^{(1)}$ in formula (\ref{Pi}). Calculating the Dirac traces and \ performing
the Wick rotation allows to explicitly determine  $\Pi$  as follows%
\begin{align}
\Pi(p_{e},m,D)  &  =-4g_{y}^{2}m^{d-2}\int\frac{dq_{e}^{d}}{(2\pi)^{d}%
}\frac{1}{1+q_{e}^{2}}+ \nonumber\\
&  g_{y}^{2}\text{ }p_{e}^{2}\int\frac{dq_{e}^{d}}{(2\pi)^{d}}\frac{1}%
{1+q_{e}^{2}}\frac{1}{1+(p_{e}-q_{e})^{2}} \nonumber\\
&  =-4g_{y}^{2}\text{ }m^{d-2}\frac{\Gamma(1-\frac{d}{2})}{(2\pi)^{\frac{d}%
{2}}}+ \nonumber\\
&  g_{y}^{2}\text{ }p_{e}^{2}\frac{m^{-2\epsilon}\Gamma(\epsilon)}%
{(4\pi)^{\frac{d}{2}}}\int_{0}^{1}dx(1+\frac{p_{e}^{2}}{m^{2}}%
x(1-x))^{-\epsilon}.%
\end{align}

The Feynman parametric integral can be explicitly performed by using the Wolfram
Mathematica program in the form %
\begin{align}
\mathcal{F(}\frac{p_{e}^{2}}{m^{2}},\epsilon\mathcal{)}  &  =\int_{0}%
^{1}dx(1+\frac{p_{e}^{2}}{m^{2}}x(1-x))^{-\epsilon}\nonumber\\
&  =-\frac{2^{-(1+\epsilon)}(q+\sqrt{q+4})(1-\frac{q}{\sqrt{q^{4}+4}%
})^{\epsilon}}{q(\epsilon-1)}\times \nonumber\\
&  _{2}F_{1}(1-\epsilon,\epsilon,2-\epsilon,\frac{1}{2}(1+\frac{q}{\sqrt
{q^{4}+4}}))- \nonumber\\
&  -\frac{2^{-(1+\epsilon)}(q-\sqrt{q+4})(1+\frac{q}{\sqrt{q^{4}+4}%
})^{\epsilon}}{q(\epsilon-1)}\times \nonumber\\
&  _{2}F_{1}(1-\epsilon,\epsilon,2-\epsilon,\frac{1}{2}(1+\frac{q}{\sqrt
{q^{4}+4}})),\\
\operatorname{Re}[\frac{\sqrt{q^{4}+4}}{q}  &  \geq1]\text{ or }%
\operatorname{Re}[\frac{\sqrt{q^{4}+4}}{q}\leq1]\text{ \ or }\frac{\sqrt
{q^{4}+4}}{q}\notin\text{real}.%
\end{align}

The exact expression for the  finite part of the one selfenergy
 can be obtained by summing the counterterms. This
 permits to define the \ wave function renormalization
constant and the coefficient of the mass counterterm as the pole part in
$\epsilon $ of the  fermion loop contribution. The wavefunction renormalization
constant is given as
\[
Z_{\phi}=1-\frac{(g_{Y}^{0})^{2}}{(4\pi)^{2}}\frac{1}{\epsilon} .%
\]

This formula  allows to find the limit  $d=4-2\,\epsilon \,-> 4$ limit of the mean of the squared
deviation of the field in (\ref{Pi}) as follows %
\begin{align}
\langle0|\alpha_{(R)}^{2}\phi_{(R)}^{r2}(x)|0\rangle &  =\frac{\alpha
_{(B)}^{2}}{Z_{\phi}}\text{ }\Pi^{(1)}(0,m,d)^{\frac{d-2}{2}}\frac
{(\pi)^{2-\epsilon}\Gamma(\epsilon-1)}{(2\pi)^{4-2\epsilon}} \nonumber\\
&  =\alpha^{2}\lim_{\epsilon\rightarrow0}\frac{\Pi^{(1)}(0,m,d)^{\frac{d-2}%
{2}}\frac{(\pi)^{2-\epsilon}\Gamma(\epsilon-1)}{(2\pi)^{4-2\epsilon}}}%
{1-\frac{(g_{Y}^{0})^{2}}{(4\pi)^{2}}\frac{1}{\epsilon}} \nonumber\\
&  =16\pi^{2}(1-\gamma)\alpha^{2}m^{2} \nonumber\\
&  =16(\frac{3}{4})^2\pi^{2}(1-\gamma)\text{ }\kappa^{2}m^{2},\\
\gamma &  =0.57721.
\end{align}

This result indicates that the mean squared deviation of the argument of the
exponential of the Dilaton radiation field  is of the order of the
square of ratio between the mass of the fermion and the Planck mass. Therefore,
for the values considered in the work: the top quark mass and the GUT unification scale, the exponential
interaction term in the Dilaton radiation field should be well approximated by
the linear term defining the Yukawa model.

\subsection{Approximations in determining Dilaton stabilization and mass }

In this subsection,  we describe the main criteria of stabilization of the vev and
the approximations done in evaluating the mass  in the paper.
Let us consider the definition of the effective action of a scalar field  \
\[
\Gamma\lbrack\phi(x)]=\frac{1}{i}\ln Z[j]-\int dx\text{ }j(x)\phi(x),
\]
and its expansion around an homogeneous vev $\phi$ in a functional power series
in spacetime dependent fluctuation $\varphi(x)$,
\begin{align}
\Gamma\lbrack\phi+\varphi(x)] &  =\Gamma\lbrack\phi]+\int dxj(x)\varphi
(x)-\frac{1}{2}\int\int dxdy\varphi(x)D^{-1}(x-y)\varphi(y)+... \nonumber\\
&  =-V^{(4)}V[\phi]+\int dp\text{ }j^{\ast}(p)\varphi(p)-\frac{V^{(4)}}{2}%
\int\int\varphi^{\ast}(p)D^{-1}(p)\varphi(p)+... \nonumber\\
&  =V^{(4)}(V[\phi]+\int dp\text{ }j^{\ast}(p)\varphi(p)-\int\int\varphi
^{\ast}(p)D^{-1}(p)\varphi(p)...),
\end{align}
where due to the homogeneity of the vev $\Gamma\lbrack\phi]$ \ is proportional
\ to the four \ dimensional volume $V^{(D)}$. The negative sign in the quadratic in the field term
 of the expansion comes because
the second functional derivative of the effective action respect to the field
is the negative of the inverse of the propagator.
 The quantity $V$ appearing  defines the so called effective potential
$V[\phi]=-\frac{\Gamma\lbrack\phi]}{V^{(4)}}$. The potential $V$\ has the
interpretation of the \ energy density of the quantum system at the value of
the also homogeneous external source that sustains it. The stable states of the vev
fields are given by the minima of $V$ \cite{jona,coleman}. \ This property implies
that the results in the work indicate the
stabilization of the Dilaton vev due to the quantum corrections associated to
massive matter fields.

\ \ \ Lets us consider now the question of the approximation in which the
results determine an estimation of the mass for the oscillation around the
mean field.\ The discussion in the previous  subsection  partially helps to clarify
this point. The masses of the particles are defined by the squared momenta
making equal to zero  the inverse propagator
\[
D^{-1}(p)=-(p^{2}-\Pi^{(1)}(p,m,4))=0.
\]
where, since $\Pi^{(1)}$ is finite,  $d=4$ has been substituted. The approximation which is
employed to estimate the Dilaton mass in the work
is obtained by expanding the proper mass (the squared momentum value) \ which
makes vanish the inverse propagator, in powers of the coupling constant
$g_{Y}$ as follows
\[
p^{2}(g_{Y})=\sum_{m=0,1,2...}p_{m}^{2}(g_{Y}^{0})^{2m}.
\]
After substituting this  series in the inverse propagator, and noting that the
inverse propagator for the scalar field is a function of the momentum only
through its squared values and that the coupling expansion of the one loop selfenergy
$\Pi^{(1)}$ is already of order  $(g_{Y}^{0})^{2}$, it follows
\begin{align}
-p_{0}^{2} &  =0, \nonumber\\
(-p_{1}^{2}(g_{Y}^{0})^{2}+\Pi(0,m,4)) &  =0, \nonumber\\
m^{2} &  =p^{2}=\Pi^{(1)}(0,m,4)+O((g_{Y}^{0})^{4}).
\end{align}

Therefore, in the first approximation, the proper mass is given by the squared
root of the zero momentum component of the selfenergy. Further,  the selfenergy
at zero momentum coincides with the squared root for the second derivative of
the effective potential, and this  was the criterion employed to evaluate the
Dilaton mass in the work. This property, can be seen after expressing the
effective action expanded around an assumed minimum and homogeneous field
value as follows%
\begin{align}
\Gamma\lbrack\phi+\varphi(x)] &  =\Gamma\lbrack\phi(x)]-\frac{1}{2}\int\int
dxdy\varphi(x)D^{-1}(x-y)\varphi(y)+... \nonumber\\
&  =-V^{(4)}V[\phi]-\frac{V^{(4)}\varphi^{2}}{2}\int\int dzD^{-1}(z)+... \nonumber\\
&  =-V^{(4)}V[\phi]-\frac{V^{(4)}\varphi^{2}}{2}[D^{-1}(p)]_{p=0}+... \nonumber\\
&  =-V^{(4)}(V[\phi]+\frac{\varphi^{2}}{2}[\Pi(p)]_{p=0}+..) \nonumber\\
&  =-V^{(4)}(V[\phi]+\frac{\varphi^{2}}{2}\frac{\partial^{2}}{\partial\phi
^{2}}V(\phi)+...,)\\
\frac{\partial^{2}}{\partial\phi^{2}}V(\phi) &  =[\Pi(p)]_{p=0}.
\end{align}


\begin{thebibliography}{99}
\bibitem{GSW}   Green M B, Schwartz J H  and Witten E, \textit{%
Superstring theory} (Cambridge University Press, Cambridge, 1987).

\bibitem{Ven} Veneziano G, \textit{Scale factor duality for classical and
quantum strings, 1991  Phys.\ Lett.\ B \textbf{265}, 287}.

\bibitem{TV}  Tseytlin A A and Vafa C, \textit{Elements of string cosmology,
1992 Nucl.\ Phys.\ B \textbf{372}, 443}, [arXiv:hep-th/9109048].

\bibitem{bound}  Adelberger E G, Heckel B R and Nelson A E, \textit{Tests of
the gravitational inverse-square law, 2003  Ann.\ Rev.\ Nucl.\ Part.\ Sci.\
\textbf{53}, 77}, [arXiv:hep-ph/0307284]. 

\bibitem{Polyakov}  Damour T and Polyakov A M, \textit{The string dilaton and
a least coupling principle, 1994 Nucl.\ Phys.\ B \textbf{423}, 532},
[arXiv:hep-th/9401069]. 


%




\bibitem{DS}  Dasgupta K , G.~Rajesh and S.~Sethi, \textit{M theory, orientifolds
and G-flux, 1999 JHEP \textbf{9908}, 023}, [arXiv:hep-th/9908088].

\bibitem{GKP}  Giddings S B, Kachru S and Polchinski J, \textit{Hierarchies from
fluxes in string compactifications, 2002 Phys.\ Rev.\ D \textbf{66}, 106006},
[arXiv:hep-th/0105097]. 

\bibitem{gaugino}  Ferrara S, Girardello L and Nilles H P, \textit{Breakdown of
local supersymmetry through gauge fermion condensates, 1983 Phys.\ Lett.\
B \textbf{125}, 457};
\newline Affleck  I, Dine M and
Seiberg N, \textit{Supersymmetry breaking by instantons, 1983 Phys.\ Rev.\
Lett.\ \textbf{51}, 1026};
\newline Affleck I, Dine M and
Seiberg N, \textit{Dynamical supersymmetry breaking in supersymmetric QCD, 1984
Nucl.\ Phys.\ B \textbf{241}, 493};
\newline Affleck I,
Dine M and Seiberg N, \textit{Dynamical supersymmetry breaking in
four-dimensions and its phenomenological implications, 1985 Nucl.\ Phys.\
B \textbf{256}, 557 };
\newline Shifman M A and
Vainshtein A I, \textit{On gluino condensation in supersymmetric gauge
theories. SU(N) and O(N) groups, 1988 Nucl.\ Phys.\ B \textbf{296}, 445}
[1987 Sov.\ Phys.\ JETP \textbf{66}, 1100].

\bibitem{heterotic}  M.~Dine, R.~Rohm, N.~Seiberg and E.~Witten, \textit{Gluino
condensation in superstring models, 1985 Phys.\ Lett.\ B \textbf{156}, 55}.

\bibitem{Danos}  Danos R J, Frey A R and Brandenberger R H, \textit{
Stabilizing moduli with thermal matter and nonperturbative effects},
[arXiv:hep-th/0802.1557]. 

\bibitem{BV}  Brandenberger R H and Vafa C, \textit{Superstrings in the early
Universe, 1989 Nucl.\ Phys.\ B \textbf{316}, 391}.

\bibitem{SGCrevs}  Brandenberger R H, \textit{String gas cosmology and structure
formation: A brief review, 2007 Mod.\ Phys.\ Lett.\ A \textbf{22}, 1875},
[arXiv:hep-th/0702001]; \newline
Brandenberger R H, \textit{Moduli stabilization in string gas cosmology, 2006 Prog.\
Theor.\ Phys.\ Suppl.\ \textbf{163}, 358}, [arXiv:hep-th/0509159];%
\newline
Battefeld T and Watson S, \textit{String gas cosmology, 2006 Rev.\ Mod.\ Phys.\ \textbf{%
78}, 435}, [arXiv:hep-th/0510022]. 

\bibitem{NBV}  Nayeri A., Brandenberger R. H.  and Vafa C., \textit{Producing a
scale-invariant spectrum of perturbations in a Hagedorn phase of string
cosmology}, [arXiv:hep-th/0511140];
\newline
Nayeri A, \textit{Inflation free, stringy generation of scale-invariant
cosmological fluctuations in D = 3 + 1 dimensions}, [arXiv:hep-th/0607073].

\bibitem{BNPV2}  Brandenberger R H, Nayeri A, Patil S P and Vafa C,
\textit{String gas cosmology and structure formation, 2007 Int.\ J.\ Mod.\ Phys.\ A
\textbf{22}, 3621}, [arXiv:hep-th/0608121].

\bibitem{BNPV1}  Brandenberger R H, Nayeri A, Patil S P and Vafa C,
\text{Tensor modes from a primordial Hagedorn phase of string cosmology, 2007 Phys.\
Rev.\ Lett.\ \textbf{98}, 231302}, [arXiv:hep-th/0604126].

\bibitem{jcap}  Cabo A and Brandenberger R H, \textit{Could fermion masses play
a role in the stabilization of the dilaton in cosmology?, 2009 JCAP \textbf{02}
015}.

\bibitem{elizalde} Elizalde E, Naftulin S and Odintsov S D, \textit{One-loop
divergence in dilaton gravitation with neutral fermions, 1994 Phys.\ Rev.\ D
\textbf{49}, 2852}.

\bibitem{muta}  Muta T, \textit{Foundations of Quantum Chromodynamics,
World Scientific Lecture Notes Vol. 5} (World Scientific Publishing Co. Pte.
Ltd., Singapore, 1987).

\bibitem{schavuor}  Y.~Schroder and A.~Vuorinen, High-precision epsilon
expansion of single-mass-scale four-loop vacuum bubbles, JHEP
\textbf{0506}, 051 (2005), arXiv:hep-ph/0503209v1.


\bibitem{jona} G. Jona-Lasinio, Nuovo Cimento 34, 1790 (1964).

\bibitem{coleman} S. Coleman and E. Weinberg, Phys. Rev. D 7, 1888, (1973).


\end{thebibliography}
\end{document}